# Octahedral and polar phase transitions in freestanding films of SrTiO$_3$

*Ludmila Leroy†, Shih-Wen Huang†\*, Chun-Chien Chiu, Sheng-Zhu Ho, Janine Dössegger, Cinthia Piamonteze, Elsa Abreu, Alessandro Bombardi, Jan-Chi Yang and Urs Staub\**

L. Leroy, S.-W. Huang, C. Piamonteze, U. Staub
Paul Scherrer Institut, Forschugnsstrasse 111, 5232 Villigen, Switzerland
\* corresponding authors
E-mail: urs.staub@psi.ch
E-mail: shih.huang@psi.ch

C. Chiu, S.-Z. Ho, J.-C. Yang
Department of Physics, National Cheng Kung University, Tainan 701, Taiwan
Center for Quantum Frontiers of Research & Technology (QFort), National Cheng Kung University, Tainan, 70101 Taiwan

J. Dössegger, E. Abreu
Institute for Quantum Electronics, ETH Zürich, Auguste-Piccard-Hof 1, 8093 Zürich, Switzerland

A. Bombardi
Diamond Light Source Ltd, Diamond House, Harwell Science & Innovation Campus, Didcot, Oxfordshire, OX11 0DE

† These authors contributed equally to this work

From extreme strain to bending, the possibilities in the manipulation of freestanding films of oxide perovskites bring a novel landscape to their properties and brings them one step closer to their application. It is therefore of great importance to fully understand the inherent properties of such films, in which dimensionality and surface effects can play a major role in defining the properties of the material's ground state. This paper reports the properties of freestanding (FS) films of the canonical oxide, SrTiO$_3$ (STO) with thicknesses 20, 30, 40 and 80 nm. We show that the relaxed ultrathin STO FS films become polar at temperatures as high as 85 K, in contrast to the quantum paraelectric behavior of bulk. Our findings are based on the softening of the ferroelectric mode towards the ferroelectric transition temperature $T_c$ and its consecutive hardening below $T_c$ with further decreasing temperature, probed with THz time domain spectroscopy in transmission mode. We find almost no thickness dependence in $T_c$. Moreover, we characterize the antiferrodistortive (AFD) phase transition in STO FS by X-ray diffraction (XRD) probing superlattice reflections characteristic for the rotation of the TiO$_6$ octahedra. Our results point to a higher phase transition temperature in comparison to bulk STO, as well as an unbalanced domain population favoring the rotation axis to be in plane. X-ray linear dichroism results further show a preferential Ti $xz/yz$ orbital occupancy at the surface, but with a complete degeneracy in the $t_{2g}$ states in the inner part of the film indicating that the AFD distortion does not strongly affect the $t_{2g}$ splitting. These findings demonstrate that STO FS films have clearly different properties then bulk.

## 1. Introduction

SrTiO$_3$ (STO) is an incipient ferroelectric perovskite, [1–5] which maintains its paraelectric properties as bulk even down to the lowest temperature due to quantum fluctuations.[5] It has become one of the most investigated materials in the form of oxide thin films and heterostructures due to its versatile response to chemical changes, isotopic substitution or strain. [6–12] Epitaxially grown strained STO films can display ferroelectricity at room temperature, [13] and doping few carriers into STO can lead to superconductivity at low temperatures. [11,14,15] Strained and strain-free films, as well as single crystals of STO are relaxor ferroelectrics and it has been demonstrated that nano-polar regions with short correlation lengths are present in strain free STO films.[16] It is well known that bulk STO displays an antiferrodistortive (AFD) structural phase transition from cubic to tetragonal at 105 K,[17] based on the rotation of the TiO$_6$ octahedra and the doubling of the primitive unit cell. The proximity to a ferroelectric transition with a drastic increase in the dielectric constant for decreasing temperatures [5,18] is directly related to the softening of the polar ferroelectric TO1 mode. [19–22] Density functional theory studies on the interplay of the AFD and ferroelectricity in STO found the AFD to compete or cooperate with the rise of the ferroelectric phase for small and large octahedral rotation angles, respectively. [23] It has also been suggested that quantum confinement in STO can shift the balance between AFD and polar modes and stabilize one of them over the other. [24]

FS films of oxide perovskites open new routes in oxide sciences due to the unprecedent possibilities of manipulation such as creating extreme strain, texture engineering and direct application in flexible electronics. [25–28] The novelty of FS films of STO raises several questions with respect to its properties. STO is a highly tunable material with a rich phase diagram, therefore benchmarking the FS film properties with respect to bulk STO is a required and important step for further applications. On the experimental side, thin FS films allow experiments in transmission geometry using electrons and electromagnetic radiation in a range that is strongly absorbed by the samples, as it is the case for soft x-rays. These transmission experiments give direct access to the intrinsic properties of the whole film, which are not accessible by pure surface sensitive techniques. This is key to understand the STO FS behavior with respect to properties of both bulk single crystals and strained thin films, with a clear focus on the AFD phase transitions and the quantum paraelectric/ferroelectric properties.

In our work we investigate the structural and electronic properties of freestanding films of STO for various thicknesses utilizing x-ray diffraction, THz time domain spectroscopy and x-ray absorption linear dichroism. We determine the AFD transition and follow the ferroelectric mode behavior as a function of temperature, and demonstrate that the properties of STO FS are remarkably distinct both from its bulk counterpart as well as from epitaxially strained films.

## 2. Results and discussion

### 2.1. Film fabrication and characterization of the crystal structure

Films of STO [001] of various thicknesses were epitaxially grown with pulse laser deposition on an STO [001] substrate coated by a sacrificial layer of $La_{0.7}Sr_{0.3}MnO_3$ of approximately 10 nm thickness. The sacrificial layer was chemically etched utilizing KI (600 µM) and HCl (50 $\mu$M) solution and the FS films were transferred to either a Si-wafer or a Cu-mesh (figure 1a). [29] In this work, STO freestanding films with [001] out-of-plane direction with 20, 30, 40 and 80 nm thickness were investigated. The crystal structure of these films was investigated using x-ray diffraction at the Material Science beamline [30,31] at the Swiss Light Source. XRD proves the high quality and crystallinity of the STO freestanding films, with clear Kiessig fringes (see figure 1b) and lattice parameters that refine within the errors to the non-strained cubic structure at room temperature. As an example, values for the refined lattice parameters of a 20 nm STO FS film at room temperature are **a** = 3.909(2) Å, **b** = 3.906(2) Å, **c** = 3.905(4) Å, $\alpha$ = 89.99(2), $\beta$ = 90.01(2), $\gamma$ = 90.00(3). Our results demonstrate that no measurable strain is present, within the precision of our measurements, and that our films are structurally relaxed.

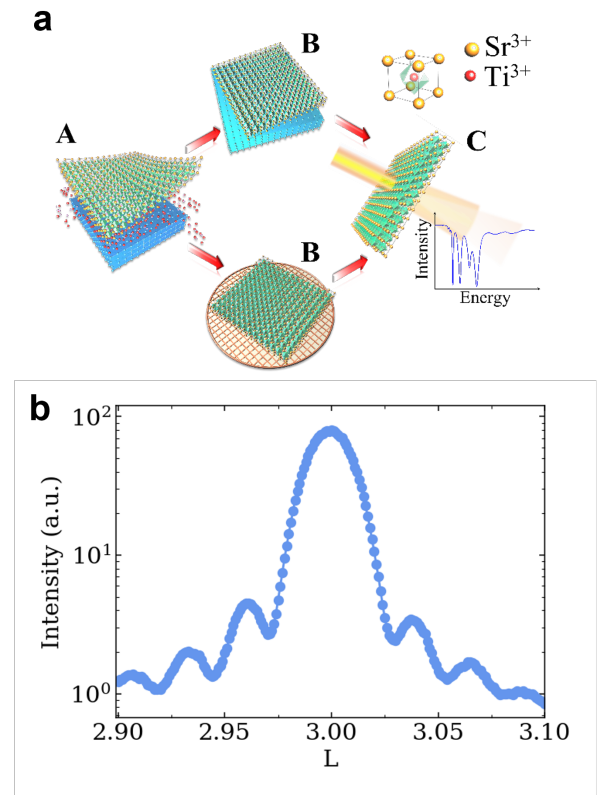

Figure 1. a) Schematic illustration of the fabrication process of FS films with the (A) wet etching of the sacrificial layer, (B) transfer of the film to an arbitrary substrate and (C) e.g. measurement of the x-ray absorption in transmission geometry. b) L scan around the (1 3 3) reflection for a 20 nm thick STO FS film deposited on a Si wafer using an x-ray energy E = 12.7 keV.

## 2.2. Octahedral rotations

As the AFD transition in STO can be influenced by strain and isovalent substitution, e.g. of Sr by Ca, [6,32,33] precise determination of the AFD transition temperature $T_{AFD}$ of the FS films is required. Temperature dependent XRD measurements of a 40 nm thick STO FS film transferred to a Si wafer piece were carried at the I16 beamline of Diamond Light Source. The half-integer superlattice reflections that allow us to monitor the occurrence of the AFD phase transition (figure 2a,b) from a cubic to a tetragonal crystal lattice [10,34] is readily observed for increasing temperatures up to the AFD transition temperature $T_{AFD}$. The integrated intensity of the half integer reflections as a function of temperature can be linearly extrapolated to extract $T_{AFD}$. The linear regressions of the peak area intensities for the (5/2 5/2 1/2) and (5/2 1/2 5/2) between 75 K and 115 K (red dots in figure 2c,d) result in $T_{AFD}$ = 116.0 ± 0.5 K and $T_{AFD}$ = 123 ± 3 K, respectively. Such an increased transition temperature compared to $T_{AFD}$ = 105 K of bulk STO [17] point to a higher structural stability of the AFD phase in STO FS, reinforcing their different properties. Moreover, the (5/2 1/2 5/2) and (5/2 5/2 1/2) reflections in cubic notation correspond to the reflections (3 2 5) and (5 0 1) in the low temperature I4/mmc space group symmetry, respectively, with the latter being forbidden. Therefore, observation of both reflections indicates the presence of AFD domains with in and out-of-plane octahedral rotations axis. The measured peak intensities need to be corrected for the probed volume, which is proportional to the spot size since the films are much thinner than the absorption length. Given by the incidence angle of 4 and 21 degrees for the (5/2 5/2 1/2) and (5/2 1/2 5/2) reflections, respectively, this results in approximately five times larger projected beam area for the former reflection. The corrected intensity probing the different domain orientation remains different by a factor of five, indicating significant preference for AFD domains with in-plane rotation axis in these very thin films. For the thinner 20 nm film (see SI) which has been measured in the same experimental configuration, the difference is even larger, and which indicates that the surface, respectively the limited thickness, prefers an inplane axis for the octahedral rotations. Note that as the reflections probe individual domains the different $T_{AFD}$, (within 2 sigma), would also indicate that the two different orientations are not only differently populated, but also have different energetics to form due to the finite size of the freestanding film.

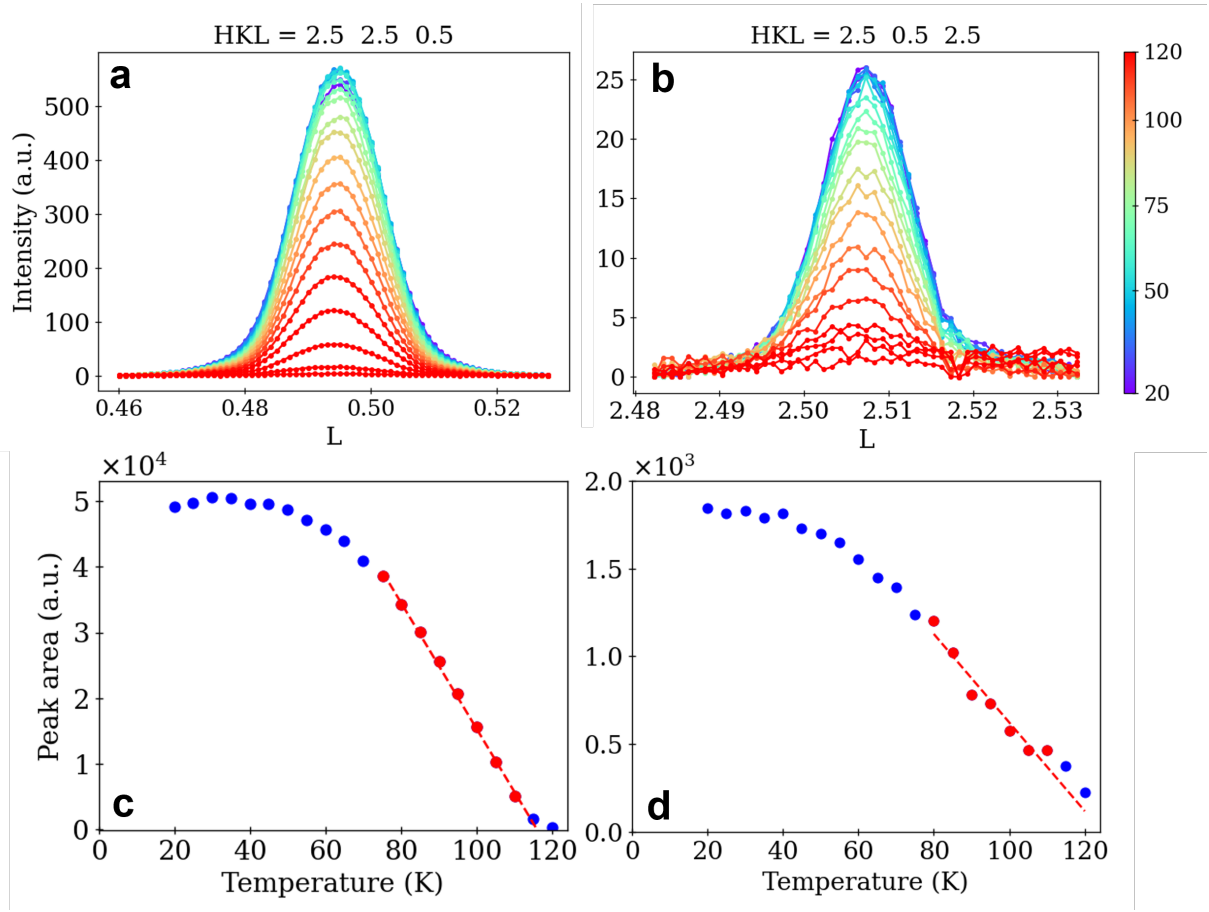

**Figure 2.** L-scans around the superlattice reflections (**a**) (5/2 5/2 1/2) and (**b**) (5/2 1/2 5/2) as a function of temperature showing their intensity profiles for the 40 nm thick STO FS. Fitted peak maximum intensity as a function of temperature for (**c**) (5/2 5/2 1/2) and (**d**) (5/2 1/2 5/2) reflections. The dotted red line shows the linear fit near the transition that results in the critical temperatures $T_{AFD}$ =116.0 $\pm$ 0.5 K and $T_{AFD}$ =123 $\pm$ 3 K respectively.

### 2.3. Ferroelectric soft mode

The concept of a polar soft mode is commonly used to explain the dynamics of phase transitions that lead to the ferroelectric state of a material. [22] This polar phonon mode describes the atomic coordinates of the motion that connects the two phases. Its frequency tends to soften when approaching the critical temperature $T_c$ of the ferroelectric phase transition. For second order phase transitions, the mode softens as the temperature decreases until its frequency goes to zero at $T_c$. As ferroelectric transitions in oxides are commonly first order, the softening is often incomplete. On strained STO thin films that are polar, the eigenfrequency of the soft-mode (the lowest optical mode) decreases with decreasing temperatures and shows a minimum at $T_c$ in which a lattice instability leads to the ferroelectric phase.[10,13] THz time domain spectroscopy (TDS) is an ideal technique to detect the infrared active soft mode frequency, and it can be done in transmission for films with tens of nm thickness, due to the strongly absorbing nature of the mode (see SI for more information). THz TDS measurements of our STO FS films with thicknesses 20, 30, 40 and 80 nm deposited on a Si wafer were performed as a function of temperature, utilizing an empty Si wafer as a reference. The Fourier transform of the time domain data from the STO FS films normalized by the reference shows the polar transverse optical mode TO1 (~2.6 THz for bulk STO at room temperature) for all sample thicknesses (see SFigure 2). Given that the THz spectrum does not give sufficient reliable data below ~0.4 THz, modeling of the response by e.g. a Drude-Lorentz model is not trivial. The frequency of the mode is therefore extracted as corresponding to the minimum in the transmission. The softening of the mode with decreasing temperature and its subsequent stiffening below $T_c$ is clearly seen in figure 3a, in which the mode frequency is plotted as a function of temperature for all film thicknesses. There is little thickness dependence on the soft mode frequency as exemplified for 100 K in Figure 3b. The minima in Fig 3a correspond to the critical temperatures at which the films become polar, and our findings show also little thickness dependence of $T_c$. To enable a more precise determination of $T_c$, we fitted a critical exponent functions (see SI for equations) for the ordered (above $T_c$) and disordered (below $T_c$) phases of STO FS (solid lines in figure 3a), which shows that the 20, 30, 40 and 80 nm thick films become polar at 85 ± 5 K, 81 ± 5 K, 87 ± 5 K and 77 ± 5 K, respectively (figure 3c), confirming no thickness dependence within the precision of our data. Moreover, TDS experiments with horizontally and vertically polarized THz for a 30 nm thick STO FS film show no polarization dependence of the THz transmission (see SI for details), which indicates a random distribution of in-plane polarized domains. The out-of-plane polarization cannot be assessed due to the reduced thickness of these thin film samples. Nevertheless, our low-temperature piezoresponse force microscopy results support the absence of a measurable out-of-plane ferroelectric polarization (see SI for detailed information).

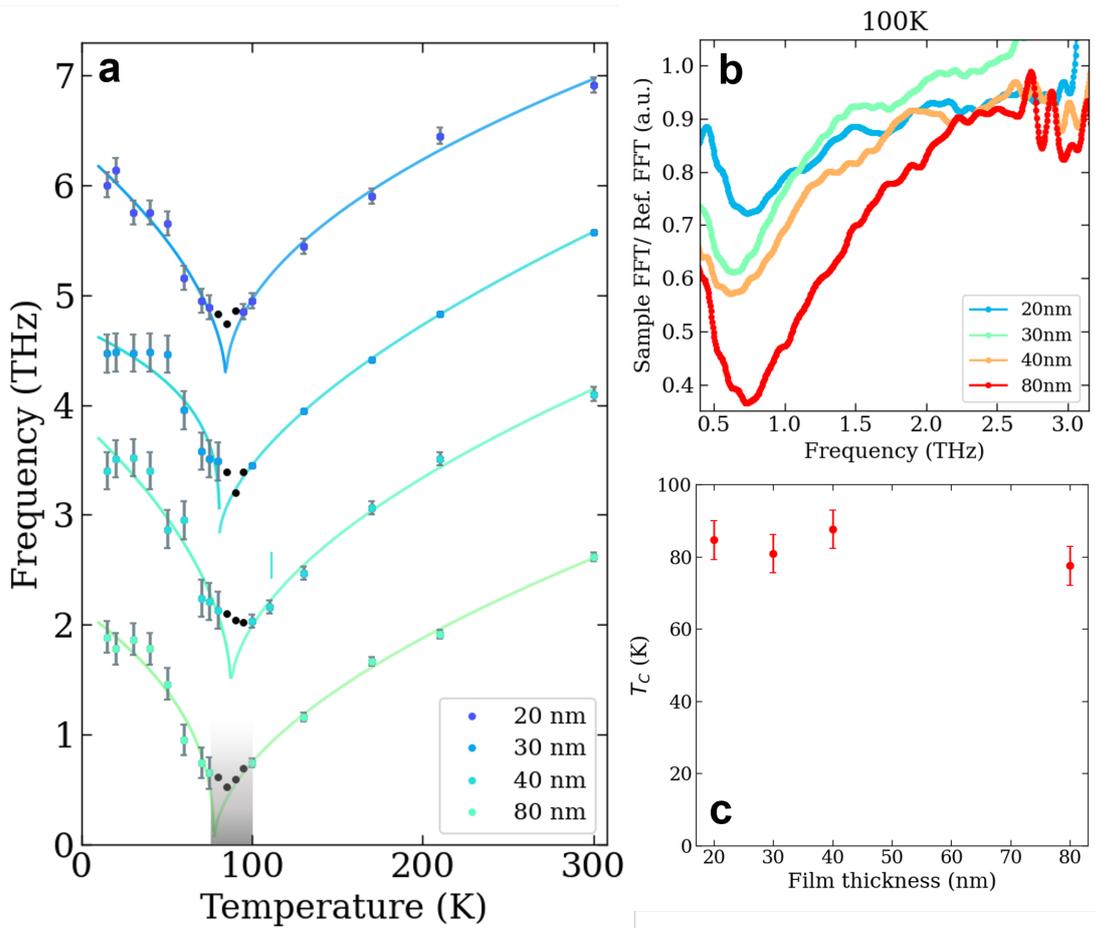

**Figure 3.** (**a**) Soft mode frequency as a function of temperature for STO FS of 20, 30, 40 and 80 nm thickness. The graphs are vertically offset by 1.4 THz sequentially starting from the dataset of the 80 nm thick film. The fits are shown in solid lines and the points in black were not included in the data fitted. The frequency range of inaccessible/imprecise values are highlighted by the shaded area. (**b**) soft mode at 100 K for all measured film thicknesses (**c**) fitted critical temperatures from the disordered (blue) and ordered (red) critical exponential fittings as a function of film thickness.

### 2.4. Anisotropy of Ti 3*d* states

One of the great advantages of ultrathin freestanding films is that they allow for transmission experiments utilizing highly absorbing x-ray energies. Therefore, the integral, intrinsic absorption can be probed directly, in contrast to total electron yield (surface sensitive) and total fluorescence yield methods, which are not direct absorption measurements. [35–37] A 60 nm thick STO [001] FS film mounted on a Cu-mesh [29] was investigated with x-ray linear dichroism (XLD) at the Ti $L_{2,3}$ edges in transmission and total electron yield modes at the XTREME beamline [38] of the Swiss Light Source. XLD with 30 degrees grazing incidence for various temperatures between 300 and 40 K have been collected. A clear but small <3% XLD signal is visible in the data collected with sample surface sensitivity (~2 nm probe depth [39,40]) in TEY mode (orange curves in figure 4b), which is roughly temperature independent (see SI). The transmitted x-rays on the other hand show almost no XLD (figure 4), highlighting the crucial difference between the TEY and transmitted components and between the surface and volume contributions in the STO FS films. The observed surface

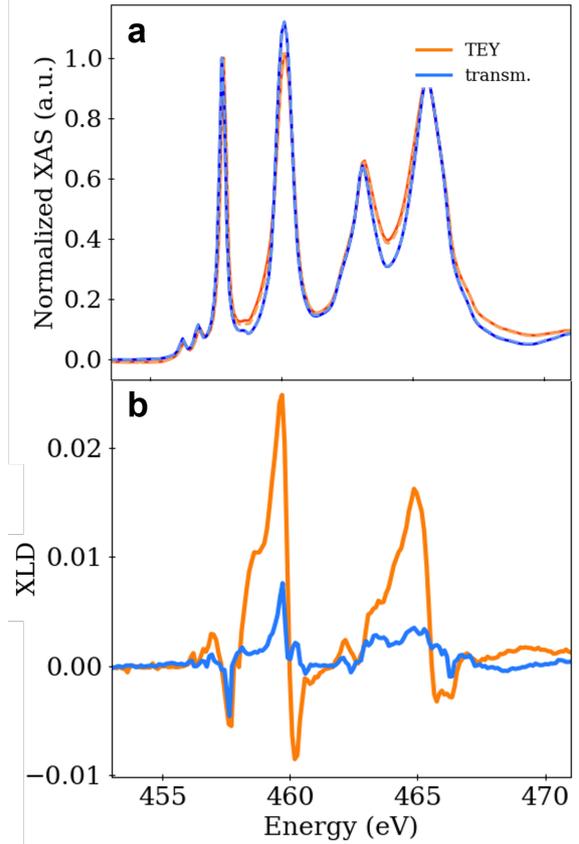

**Figure 4. a)** XAS spectra taken at 30 degrees incidence of 60 nm thick STO FS film at 100 K in transmission (blue) and TEY (orange) modes for both π (solid lines) and σ (dashed lines) polarizations. **b)** Average x-ray linear dichroism ($I_\pi$-$I_\sigma$) of STO FS with 30 nm thickness deposited on a Cu grid. Data are taken in TEY (orange) and in transmission mode (blue).

XLD is commonly observed at STO surfaces and interfaces in which the translational symmetry of the lattice is broken and the surface/interface 2D character leads to splitting of the $t_{2g}$ states. [41–43] The features of the grazing incidence XLD in TEY arise from the unoccupied Ti $d_{xy}$ states and Ti $d_{xz/yz}$ and $d_{3z^2-r}$ states, probed by π and σ polarized x-rays at gracing incidence, respectively. The observed XLD agrees with calculations [43] for the case in which $d_{xz/yz}$ is the preferential state of interfacial electrons $(d_{xy} > d_{xz/yz})$ as observed in AlO/SrTiO3/NdGaO3 interface, [43] however, the observed amplitude is much smaller and resemble more the one measured on bare SrTiO3[41] or on γ-Al2O3/SrTiO3.[44] Note that the ground state of the surface orbitals is affected by the termination of the surface, the chemical composition as well as by the surface steps. [45,46]. As the XLD taken in transmission is approximately temperature independent, we can take an average over the transmission XLDs for the obtained temperatures, which results in a XLD signal that is mat most 1/4 of the TEY signal. This indicates that the residual XLD is mostly coming from the surface and that the inner part of the film is isotropic within the probed volume. This shows that there is little splitting of the $t_{2g}$ states in the volume despite the unequal AFD domain population and the preferential in-plane polar domains.

## 3. Conclusion

In our work we investigated the canonic oxide perovskite, STO, in its freestanding film form. We show the high crystalline quality of our STO FS films and show that the films are structurally relaxed to the cubic STO bulk structure. We determined the AFD transition temperature on STO FS films with the rotation of the TiO6 octahedra occurring at a slightly higher temperature when compared to bulk. The observed Bragg peaks arising with the tetragonal symmetry support an unequal distribution of in-plane and out-of-plane rotation axis domains within the STO FS, with a preference to having an in-plane rotation axis. We also showed that the ferroelectric soft mode in STO FS softens as a function of temperature and that the films become polar near 82 K. The consecutive hardening of the mode as temperature continues to decrease demonstrates the ordered nature of the ferroelectric phase. Moreover, the critical temperatures $T_c$ do not drastically depend on the film thicknesses for films as thick as 80 nm and the ferroelectric mode shows no in-plane anisotropy of the polar domains. Finally, the XLD in TEY and transmission modes benchmark the differences between surface and bulk, with the clear indication that the surface has $d_{xz/yz}$ character, without a preferred orientation of states in the inner part of the thin films.

## 4. Experimental Section/Methods

*X-ray diffraction:* Structural characterization of all STO FS films were done by x-ray diffraction at the Material Science Beamline at the Swiss Light Source, [30,31] with an x-ray energy E = 12.7 keV and 500 x 500 $\mu$m beam size. Lattice parameters at room temperature were thoroughly benchmarked with a series of measurements of structure Bragg peaks for the 20 nm thick STO FS deposited on a Si-wafer piece. Various combinations of peak positions were used to refine the lattice parameters, which are found to be cubic at room temperature.

*Temperature dependent x-ray diffraction*: XRD measurements of the superlattice reflections of STO FS [001] of 20 and 40 nm thicknesses were performed at the i16 beamline at Diamond light source. [47] The x-ray energy was set to 11 keV with a beam size of 180 $\mu$m x 30 $\mu$m, and temperatures ranged from 120 to 20 K.

*THz Time domain spectroscopy*: THz-TDS experiments for STO freestanding films of thicknesses 20, 30, 40 and 80 nm deposited on a silicon wafer piece were performed in transmission mode at the Institute of Quantum Electronics at the ETH. We used a standard THz-TDS setup that allows the measurement of the transmission in the frequency range of ~0.5–3.0 THz. [48] A Si wafer, identical to the one on which the sample was transferred, was used as a reference.

*Piezoresponse force microscopy:* Piezoresponse force microscopy hysteresis measurements were acquired using a commercial cryogenic scanning probe microscope system (attoAFM I, Attocube) with a closed-cycle cryostat (attoDRY 2100 with 9 T magnet, Attocube) at 1.6 K and 70K. The switching spectroscopic technique was carried out under contact-resonance mode with the commercial platinum silicide (PtSi) coated tips with spring constant of 2.8 N/m (NANOSENSORS PtSi-FM). The tip was driven with an AC voltage amplitude of about 2 V and was working at a contact-resonance frequency of about 300 kHz. The off-field hysteresis loops were obtained via the switching spectroscopic technique with an arbitrary waveform generator (G5100A, Picotest).

*X-ray linear dichroism:* X-ray absorption spectra of a 60 nm thick STO FS film were measured as a function of temperature in both transmission and total electron yield modes at the X-TREME beamline [38] at the Swiss Light Source. The measurements were performed with 30-degree grazing incidence x-rays with $\pi$ and $\sigma$ polarization and a (250 $\mu m$ x 30 $\mu m$) beam size probing mainly out-of-plane and in-plane directions, respectively. XLD was corrected for tiny energy shifts between the spectra in different x-ray polarizations at normal incidence. Energy dependent data in both polarizations was acquired and normalized by the average of the XAS maxima (XAS calculated as the average of both polarizations). XLD for each temperature consists of $(I_\pi - I_\sigma)$.

**Supporting Information**

Supporting Information is available from the Wiley Online Library or from the author.

**Acknowledgements**

L. L. acknowledges the National Centers of Competence in Research in Molecular Ultrafast Science and Technology (NCCR MUST-No. 51NF40-183615) from the Swiss National Science Foundation. .-C.Y. acknowledged the financial support from National Science and Technology Council (NSTC) in Taiwan, under grant no. NSTC 112-2112-M-006-020-MY3. The authors also thank MSSORPS Co., Ltd. for high-quality TEM sample preparation and preliminary examination. The research was also supported in part by Higher Education Sprout Project, Center for Quantum Frontiers of Research & Technology (QFort) at National Cheng Kung University, Taiwan. E.A. and J.D. acknowledge support from the Swiss National Foundation through Ambizione Grant PZ00P2_179691 and E.A. through Starting Grant TMSGI2_211211.